\documentclass{article}
\usepackage{arxiv}
\usepackage[authoryear]{natbib}
\usepackage[utf8]{inputenc} % allow utf-8 input
\usepackage[T1]{fontenc}    % use 8-bit T1 fonts
\usepackage{hyperref}       % hyperlinks
\usepackage{url}            % simple URL typesetting
\usepackage{booktabs}       % professional-quality tables
\usepackage{amsfonts}       % blackboard math symbols
\usepackage{nicefrac}       % compact symbols for 1/2, etc.
\usepackage{microtype}      % microtypography
\usepackage{graphicx}
\usepackage{subcaption}
\usepackage{tcolorbox}
\usepackage{multirow}
\usepackage{diagbox}
\usepackage[normalem]{ulem}
\usepackage{doi}
\usepackage{placeins} % 引入 placeins 宏包
\usepackage{enumitem}
\usepackage{amsmath} 
\usepackage{tabularx} % 在导言区导入 tabularx 包
\usepackage{hyperref}
\usepackage{makecell}
\usepackage{tcolorbox}
\tcbuselibrary{listings}
\usepackage{algorithm}
\usepackage[noend]{algpseudocode}
\newtcblisting{mylatexcode}{
	listing only,
	colback=gray!5!white,  % 背景色
	colframe=gray!50!black, % 边框颜色
	listing options={basicstyle=\ttfamily\small}
}
\newcommand{\design}{FlexLink}

\title{\design{}: Boosting your NVLink Bandwidth by 27\% without accuracy concern.}
\author{Ao Shen, Rui Zhang, Junping Zhao \\ 
Asystem@Ant Group}
\begin{document}
	\maketitle
	\thispagestyle{firstpage}
	\begin{abstract}
		As large language models (LLMs) continue to scale, multi-node deployment has become a necessity. Consequently, communication has become a critical performance bottleneck. Current intra-node communication libraries, like NCCL, typically make use of a single interconnect such as NVLink. This approach creates performance ceilings, especially on hardware like the H800 GPU where the primary interconnect's bandwidth can become a bottleneck, and leaves other hardware resources like PCIe and Remote Direct Memory Access (RDMA)-capable Network Interface Cards (NICs) largely idle during intensive workloads. We propose \design{}, the first collective communication framework to the best of our knowledge designed to systematically address this by aggregating these heterogeneous links—NVLink, PCIe, and RDMA NICs—into a single, high-performance communication fabric. \design{} employs an effective two-stage adaptive load balancing strategy that dynamically partitions communication traffic across all available links, ensuring that faster interconnects are not throttled by slower ones. On an 8-GPU H800 server, our design improves the bandwidth of collective operators such as AllReduce and AllGather by up to 26\% and 27\% over the NCCL baseline, respectively. This gain is achieved by offloading 2–22\% of the total communication traffic to the previously underutilized PCIe and RDMA NICs. \design{} provides these improvements as a lossless, drop-in replacement compatible with the NCCL API, ensuring easy adoption.

		\FloatBarrier % 插入屏障，强制处理之前的浮动体

	\end{abstract}

\begin{figure}[h!]
    \centering

    % 第一个图
    \begin{minipage}[t]{0.48\textwidth}
        \centering
        \includegraphics[width=\linewidth]{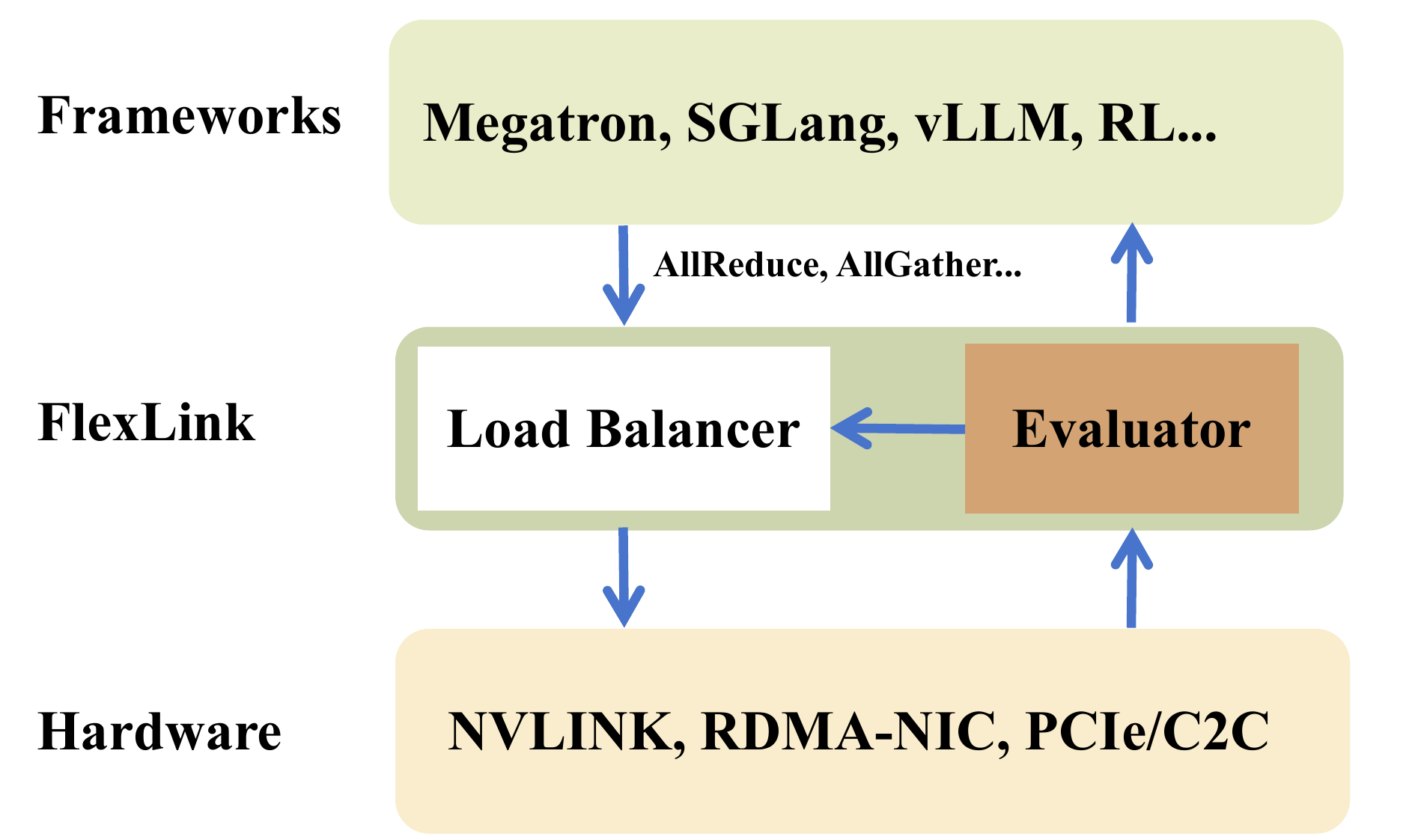}
        \caption{The design of \design{}.}
        \label{fig:design}
    \end{minipage}%
    \hfill
    % 第二个图
    \begin{minipage}[t]{0.48\textwidth}
        \centering
        \includegraphics[width=\linewidth]{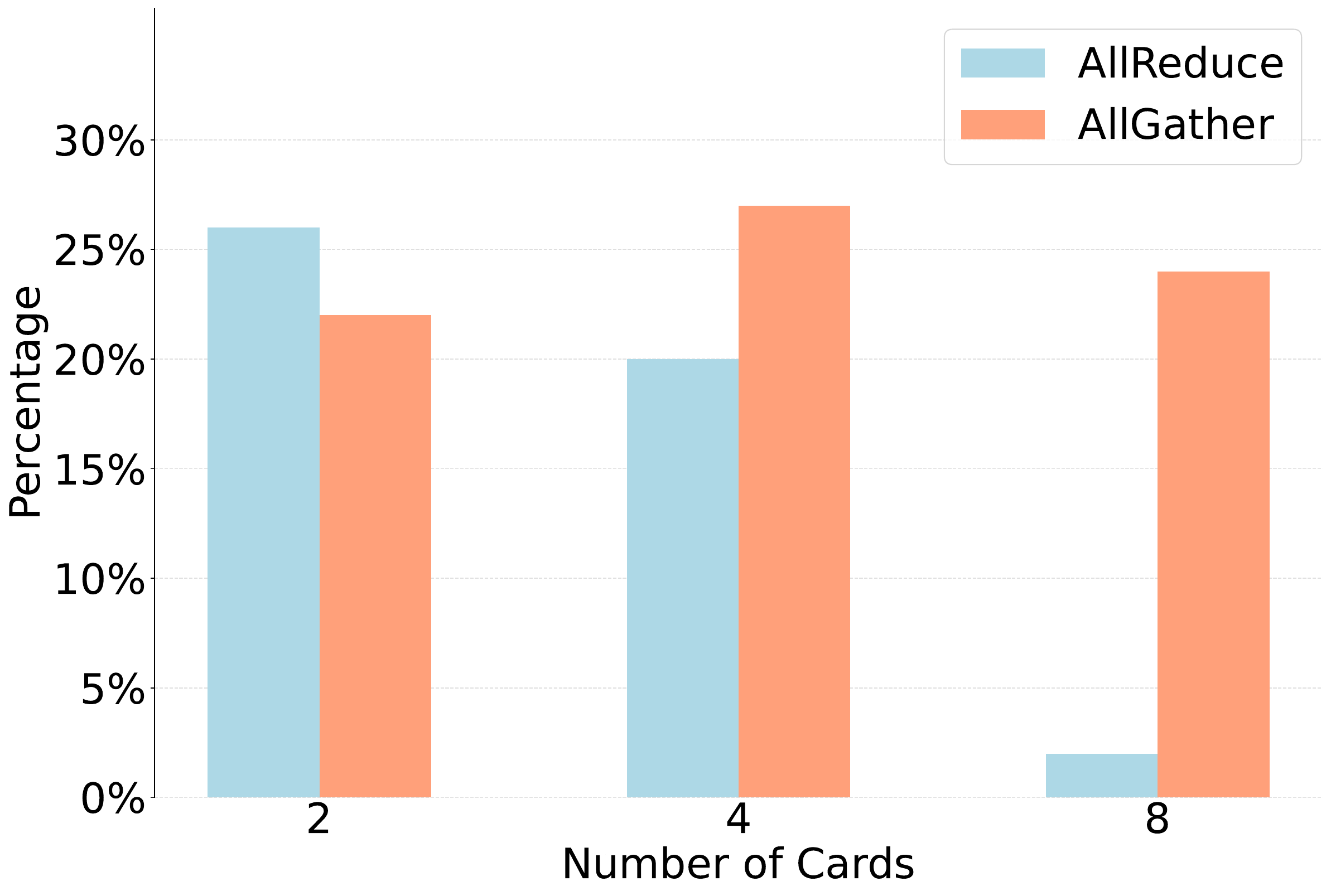}
        \caption{Bandwidth improvement of \design{} over NCCL for a 256MB message size.}
        \label{fig:result}
    \end{minipage}

\end{figure}

\section{Introduction} \label{sec:intro}
The scale of large language models (LLMs) has grown at a breathtaking pace, with models comprising hundreds of billions or even trillions of parameters becoming increasingly common~\cite{deepseekai2025deepseekv3technicalreport, qwen2025qwen25technicalreport, bai2023qwentechnicalreport}. This remarkable growth, often referred to as "scaling," has been the primary driver behind recent breakthroughs in artificial intelligence, powering transformative applications from deep research~\cite{zheng2025deepresearcherscalingdeepresearch} to advanced scientific discovery. However, this progress comes at a significant cost. Building and operating the massive AI infrastructure required for these models incurs substantial capital and operational expenditure. Consequently, improving the efficiency of these systems is not merely an optimization but a critical necessity for sustaining future innovation. Among the various factors influencing system efficiency, communication has emerged as a dominant and persistent bottleneck in both large-scale training and inference workflows.

The severity of this communication bottleneck is evident across critical AI workflows. For instance, during \textbf{Mixture-of-Experts (MoE) model training}, the intense communication required to manage vast parameter spaces can consume as much as 43.6\% of the forward pass time~\cite{jin2025megascalemoe}, severely capping efficiency. The problem is particularly acute in \textbf{long-sequence inference}, where communication overhead in Flash Communication~\cite{li2024flashcommunicationreducingtensor} can be up to 65.9\%. Moreover, our empirical analysis of a 32B model on a standard 8-H800 setup shows that for a 64K sequence length, communication during the prefill stage accounts for a significant 36\% of the total execution time

% 子问题场景引入 (约0.25页)
To handle these demanding communication patterns, modern GPU servers are equipped with a sophisticated hierarchy of interconnects: (1) NVLink, a high-bandwidth, low-latency interconnect for direct GPU-to-GPU communication within a node; (2) PCIe (Peripheral Component Interconnect Express), a bus that connects GPUs to the host CPU and, by extension, to each other via host memory; and (3) Remote Direct Memory Access (RDMA)-capable Network Interface Cards (NICs), which enable high-speed, low-overhead communication between nodes across a network. Together, these interconnects form a powerful, albeit complex, communication substrate.

% 现有工作问题 (约0.25页)
However, existing collective communication libraries (CCLs), which are fundamental to distributed ML workloads, fail to exploit the full potential of this hardware substrate. Libraries like NVIDIA Collective Communication Library (NCCL)~\cite{ncclgithub}, the de facto standard, are highly optimized but adopt a rigid communication strategy. For intra-node communication, NCCL almost exclusively utilizes NVLink when available, treating it as the sole high-performance path. While effective, this design choice leaves the considerable bandwidth of the PCIe bus and the intra-node capabilities of RDMA NICs completely idle during these operations. This underutilization of available hardware resources artificially constrains the total achievable communication bandwidth within a node. This problem is further exacerbated by compliance-driven hardware limitations. For instance, the widely deployed H800 GPU, a variant designed for compliance in certain markets, features a significantly curtailed NVLink bandwidth of 400 GB/s—less than half of the standard H100's 900 GB/s. This hardware downgrade places an even greater strain on the already overburdened primary communication path, making the underutilization of other interconnects a critical performance liability.

% Strawman方案、挑战 (约0.25页)
Addressing this underutilization is not low-hanging fruit and presents significant systems-level challenges. A strawman solution would be to simply enable these idle links and use them in parallel with NVLink. The various interconnects exhibit highly heterogeneous performance characteristics in terms of bandwidth and latency. A naive, static partitioning of data across these links would inevitably lead to the faster link (NVLink) being throttled by the slower ones (PCIe, RDMA), potentially degrading performance rather than improving it. An effective solution requires an adaptive load balancing mechanism that can dynamically adapt to the runtime state of each link and the specific demands of the communication workload. 

% 我们的设计和效果 (约0.25页)
We propose \design{}, the first collective communication framework that dynamically aggregates heterogeneous interconnects (such as NVLink, PCIe, and RDMA) into a unified, high-performance communication fabric to maximize intra-node throughput. To intelligently partition and schedule traffic across these links, \design{} employs a two-stage adaptive load balancing strategy. In the first stage, the \textit{Communicator} establishes an initial traffic partitioning plan during initialization. In the second stage, a runtime \textit{Evaluator} continuously monitors link performance, providing feedback to a \textit{Load Balancer} that dynamically refines the traffic distribution to ensure optimal resource utilization. Our evaluation on an 8-GPU H800 server demonstrates that for bandwidth-bound collective operations, \design{} consistently outperforms the highly-optimized NCCL baseline across a range of data sizes. Specifically, \design{} achieves a performance improvement of up to 26\% and 27\% for AllReduce and AllGather, respectively. We present sample results in Figure~\ref{fig:result}. Importantly, \design{} is engineered as a lossless enhancement that maintains full compatibility with NCCL API, allowing developers to harness its benefits without any modifications to their application code.

We summarize our main contributions as follows:
\begin{itemize}
     \item We design and implement \design{}, to the best of our knowledge, the first communication framework to systematically aggregate heterogeneous intra-node interconnects into a unified, high-performance communication fabric.
    \item We develop a novel two-stage adaptive load balancing strategy that ensures superior performance over NCCL under diverse communication demands.
    \item We deliver \design{} as a lossless, easy-to-use solution by maintaining compatibility with NCCL API, enabling seamless integration into existing systems.
    \item We conduct a comprehensive evaluation demonstrating that \design{} significantly outperforms the state-of-the-art NCCL, improving the communication bandwidth on an 8-GPU H800 server for AllReduce and AllGather by up to 26\% and 27\%, respectively.
\end{itemize}

	\section{Background and Motivation}
\subsection{Background}
\label{subsec:background}
Collective communication is a cornerstone of distributed deep learning, facilitating the complex data exchanges required for parallel training and inference across multiple GPU accelerators. To optimize these operations, several specialized communication libraries have been developed.

% Baseline
\textbf{NCCL.} NCCL has become the de facto standard for NVIDIA GPUs. It provides highly optimized, topology-aware implementations of collective communication primitives. For intra-node communication, NCCL typically prioritizes the highest-bandwidth interconnect available, which is usually NVLink. In the absence of NVLink, it can utilize PCIe for peer-to-peer communication. While effective, this "winner-takes-all" strategy often leads to the underutilization of other available interconnects, such as the PCIe bus. This design choice, which favors a single transport path, misses opportunities for bandwidth aggregation, especially on modern servers equipped with multiple types of high-speed links.

% Feature that we leverage
\textbf{NVSHMEM.} NVSHMEM~\cite{nvshmemnvidia} implements the OpenSHMEM parallel programming model based on Partitioned Global Address Space (PGAS). It enables fine-grained, low-latency, GPU-initiated communication through one-sided operations such as "put" or "get". Combined with technologies like InfiniBand GPUDirect Async (IBGDA),  it allows GPUs to control communication and bypass the CPU. While powerful for specific communication patterns, NVSHMEM is not a collective communication library itself. Our collective primitives for the RDMA NIC path are built with NVSHMEM and a lightweight synchronization mechanism. 

\subsection{Motivation}
\label{subsec:motivation}
Despite the proliferation of high-speed interconnects in modern servers, we observe that a significant portion of the available communication bandwidth remains untapped, creating performance bottlenecks for communication-intensive workloads.

\subsubsection{Observation: Link Idleness in AI Workflows}

We identify a critical inefficiency in prevalent communication libraries: secondary interconnects like \textbf{PCIe} often remain idle while the primary \textbf{NVLink} path is saturated during collective operations. This underutilization presents a significant performance bottleneck. We highlight this issue through two communication-intensive scenarios:

\begin{figure}
    \centering

    % 第一个图
    \begin{minipage}[b]{0.48\textwidth}
        \centering
        \includegraphics[width=\linewidth]{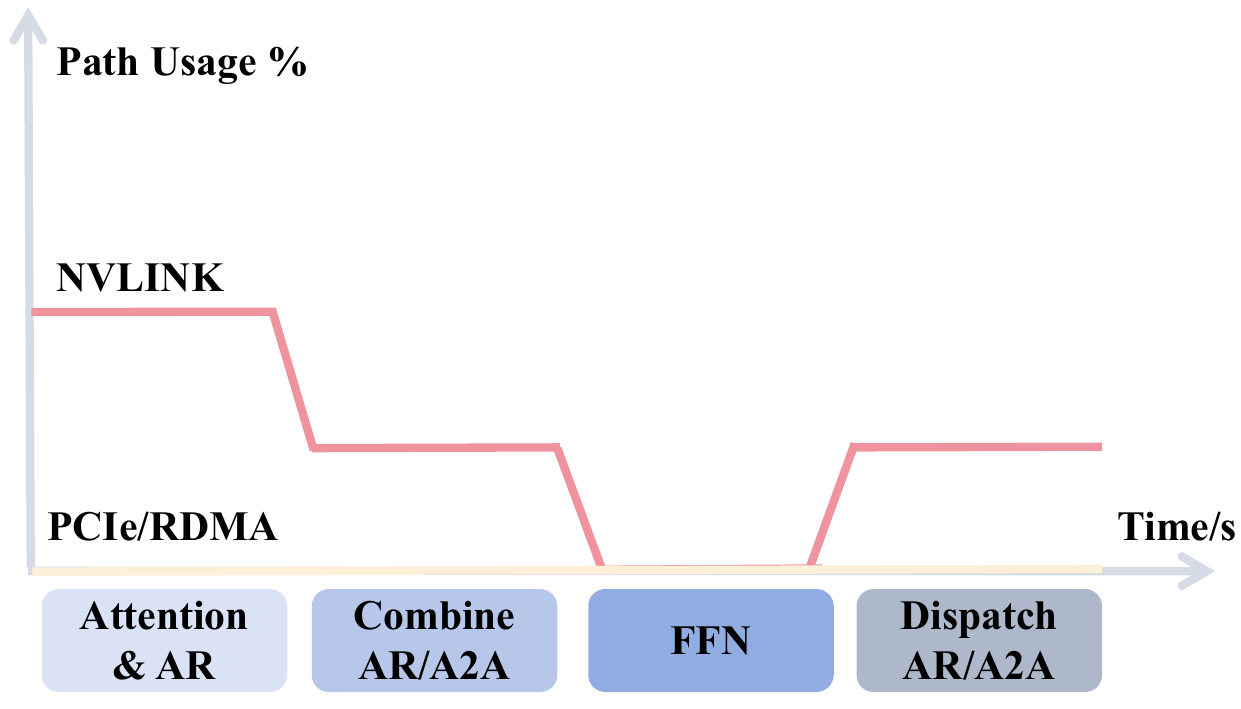}
        \caption{MoE training: Intra-node Expert (EP8) \& Tensor Parallelism (TP) with inter-node Pipeline Parallelism (PP).}
        \label{fig:design1}
    \end{minipage}% <--- 注意这个关键的注释符 %
    \hfill % 这个命令之前不应该有任何空格或换行
    % 第二个图
    \begin{minipage}[b]{0.48\textwidth}
        \centering
        \includegraphics[width=\linewidth]{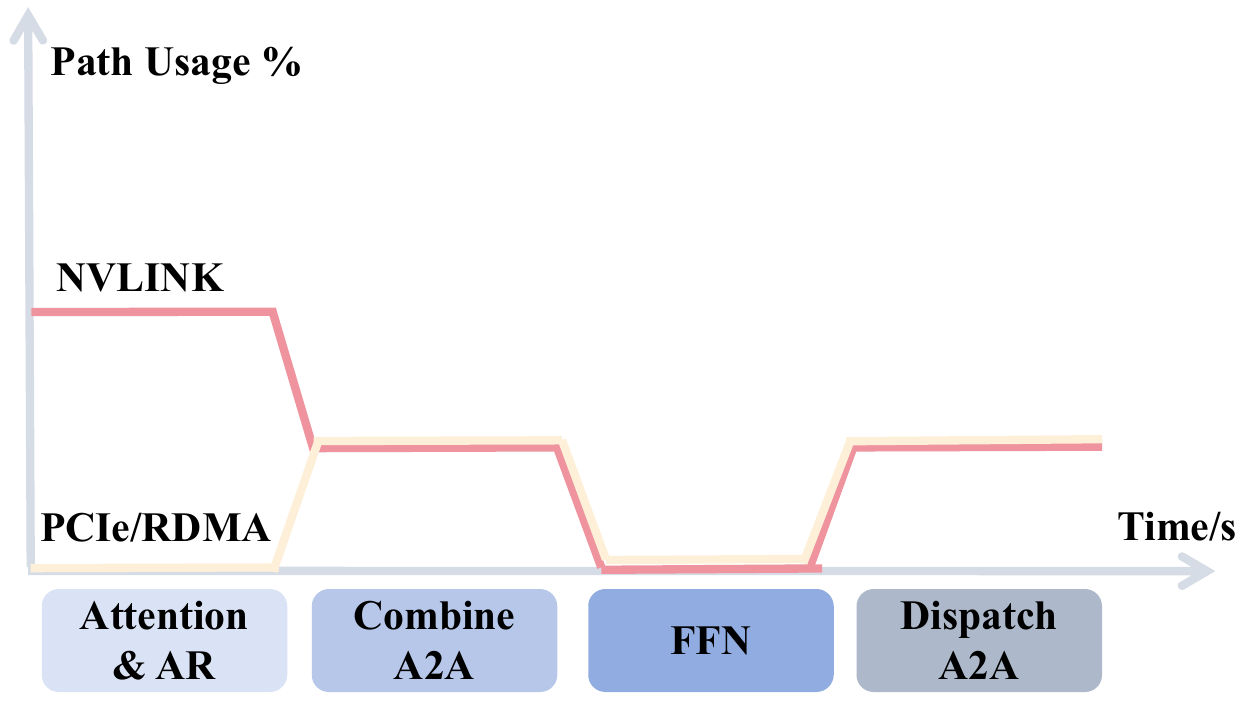}
        \caption{MoE inference: Intra-node Tensor (TP2) \& Data Parallelism (DP4) with inter-node Expert Parallelism (EP64).}
        \label{fig:design2}
    \end{minipage}

\end{figure}

\begin{itemize}
    \item \textbf{MoE model training.} 
As illustrated in Figure~\ref{fig:design1}, standard MoE training workflows make use of communication libraries like NCCL that exclusively utilize NVLink for collectives (\textit{e.g.}, AllReduce, AllToAll). Consequently, the PCIe/RDMA interconnect remains entirely idle. This becomes a critical bottleneck as MoE communication overhead can consume up to 43.6\% of the forward pass time~\cite{jin2025megascalemoe}. Harnessing the idle PCIe path thus presents a significant opportunity to alleviate this overhead and accelerate training.

    \item \textbf{Long-sequence model inference.}
This pattern of suboptimal interconnect utilization also creates a significant bottleneck in long-sequence inference, where communication overhead measured in Flash Communication~\cite{li2024flashcommunicationreducingtensor} can be up to 65.9\%. Similarly, our empirical analysis of a 32B model prefilling a 64K sequence shows that communication still accounts for a staggering 36\% of the total execution time. As illustrated in the initial attention phase of Figure~\ref{fig:design2}, this overhead is exacerbated by AllReduce operations that saturate the NVLink interconnect, while the system's PCIe links remain substantially underutilized.
\end{itemize}

\subsubsection{Observation: Link Idleness in Hardware}
\label{subsec:link_idleness}
Many GPU platforms suffer from link idleness, where secondary interconnects like PCIe and RDMA are underutilized while the primary NVLink is saturated—a problem especially severe on bandwidth-restricted GPUs like the H800. As quantified in Table~\ref{tab:idle_links}, this idle bandwidth represents a significant performance opportunity. It is important to note that the NIC configurations listed in Table~\ref{tab:idle_links} represent typical server setups; other configurations are also possible. This issue is exacerbated by path contention in current hardware. The contention arises because both GPU-to-NIC traffic (GPU $\rightarrow$ PCIe Switch $\rightarrow$ NIC) and GPU-to-CPU traffic (GPU $\rightarrow$ PCIe Switch $\rightarrow$ CPU) must traverse the same initial PCIe link connecting the GPU to the same switch. Thus, the theoretical upper limit for combined traffic over PCIe and the RDMA NIC is simply the bandwidth of the GPU's own PCIe interface (e.g., 128 GB/s for a PCIe Gen5 x16 link in H800 server).

This architectural bottleneck, however, will be resolved in future platforms. Architectures like the GB300 will decouple these I/O paths, eliminating the contention. This shift further increases the available idle bandwidth, making a solution like \design{}—which aggregates disparate links—even more critical for maximizing future hardware performance.

\begin{table}[htbp]
\centering
\caption{Analysis of Idle Bandwidth Opportunity Across GPU Architectures. 
The "Idle BW Opportunity" quantifies untapped bandwidth relative to the primary NVLink path, calculated as the ratio of total available idle bandwidth to NVLink bandwidth. 
On current platforms with path contention, the idle bandwidth is limited to the PCIe/C2C link. On future platforms without contention, it is the sum of the PCIe/C2C and RDMA NIC bandwidths.
C2C (Chip-to-Chip) refers to the GPU-CPU interconnect in the Blackwell~\cite{gb200} architecture. 
All bandwidth figures are bidirectional.}
\label{tab:idle_links}
\begin{tabular}{lccccc}
\toprule
\textbf{GPU Server} & \textbf{NVLink} & \textbf{PCIe/C2C} & \textbf{RDMA NIC} & \textbf{Path} & \textbf{Idle BW} \\
& \textbf{(GB/s)} & \textbf{(GB/s)} & \textbf{(Gb/s)} & \textbf{Contention} & \textbf{Opportunity} \\
\midrule
\multicolumn{6}{c}{\textit{Current Architectures (Shared CPU-GPU/GPU-NIC PCIe Path)}} \\
\midrule
H800 & 400 & 128 & 800 & Yes & 32\% \\
H100 / H200 / H20 & 900 & 128 & 800 & Yes & 14\% \\
A800 & 400 & 64 & 400 & Yes & 16\% \\
GB200 & 1800 & 400 & 1600 & Yes & 22\% \\
\midrule
\multicolumn{6}{c}{\textit{Next-Generation Architecture}} \\
\midrule
GB300 & 1800 & 400 & 1600 & No & 33\% \\
\bottomrule
\end{tabular}
\end{table}

\subsubsection{Observation: Inefficiency of Intra-node PCIe Communication}

Achieving PCIe theoretical bandwidth during intra-node collective operations is challenging. High software overheads and pipeline scheduling gaps prevent a single communication stream, such as a single ring, from fully saturating the physical link. A seemingly straightforward solution—employing multiple parallel rings over PCIe to aggregate bandwidth—is also largely ineffective. Our investigation reveals that concurrent, unidirectional transfers using this method are often serialized at a low level within the CUDA driver. This prevents true parallelism and fails to improve total bandwidth.

This driver-level bottleneck necessitates leveraging a logically distinct secondary communication path to fill the bandwidth gaps. The ideal candidate is the GPU-attached RDMA NIC. Although it shares the underlying PCIe switch hardware, it represents a separate endpoint that an advanced communication scheduler can address in parallel. Our experiments validate this strategy, showing that co-scheduling traffic across both the primary PCIe and secondary RDMA paths effectively utilizes otherwise idle bandwidth compared to the PCIe-only design.

% PCIe不好用
\textbf{Challenge: efficient aggregation of heterogeneous links.}
\label{subsec:challenge}
A key challenge lies in efficiently aggregating the primary NVLink interconnect with the secondary PCIe bus for intra-node communication. While communication libraries like NCCL default to using PCIe in the absence of NVLink, they do not automatically harness both links in parallel, leaving substantial PCIe bandwidth idle.

A naive attempt to force PCIe usage might involve staging data through CPU-managed host memory. This approach, however, is fundamentally flawed. The required data path (GPU $\to$ Host Memory $\to$ GPU) introduces prohibitive latency and overhead from memory copies and active CPU management. As a result, this method fails to effectively saturate the PCIe link and is impractical for performance-sensitive workloads.

\subsection{Related Works}
\textbf{Compression.}
Recent efforts have focused on compressing communication volume to mitigate network bottlenecks. Flash Communication~\cite{li2024flashcommunicationreducingtensor} introduces a low-bit compression scheme to reduce overhead in tensor parallelism for LLM inference. An evolution of this, FlashCommunication V2~\cite{li2025flashcommunicationv2bitsplitting}, enables more flexible communication at arbitrary bit widths by using bit splitting to adapt to hardware and spike reserving to handle numerical outliers during aggressive quantization. Similarly, industry frameworks like NVIDIA's TensorRT-LLM apply this principle by using lower precision formats, such as FP4, for communication primitives like AllGather~\cite{tensorrtgithub} to reduce the total data transferred over the network without impacting final accuracy.
These lossy compression methods are orthogonal to our work and thus can be combined with \design{}, our lossless bandwidth enhancement approach, to further maximize intra-node communication efficiency.

\textbf{Overlap.}
To mitigate the performance impact of communication, another line of research focuses on overlapping computation and communication kernels. Frameworks like FlashOverlap~\cite{hong2025flashoverlaplightweightdesignefficiently} propose lightweight, tile-wise signaling mechanisms to trigger communication alongside computation, which effectively reduces the invocation overhead of communication. For specific architectures, Comet~\cite{zhang2025cometfinegrainedcomputationcommunicationoverlapping} designs a fine-grained scheduling strategy to hide the extensive communication latency inherent in MoE models during training. Similarly, ConCCL~\cite{agrawal2025conccl} leverages dedicated GPU DMA engines on AMD hardware to enable concurrent execution for collective communication operations, thereby avoiding contention for the compute units.

Some recent approaches aim to automate this process through compilers. TileLink~\cite{zheng2025tilelinkgeneratingefficientcomputecommunication} introduces a compiler that automatically generates fused kernels, combining computation with communication primitives using a tile-centric model.
Our work is orthogonal to the high-level scheduling logic of these approaches. For those that exploit NCCL-based APIs, \design{} can be integrated as a more performant communication backend to further enhance their overall effectiveness.

\textbf{Inter-node multi-path.}
Recent work has explored multi-path networking to maximize communication bandwidth. FuseLink~\cite{ren2025enabling} utilizes idle GPUs to relay traffic to multiple NICs within a server, and Nezha~\cite{yu2024fullstackallreducemultirailnetworks} focuses on inter-server multi-rail networks, providing a full-stack system to schedule and load-balance traffic across multiple, potentially heterogeneous NICs (e.g., TCP, RDMA). While these systems focus on optimizing inter-server communication by leveraging multiple paths, our work, \design{}, targets the distinct problem of accelerating intra-server communication.

% In summary, \design{} uniquely provides a lossless, non-intrusive solution for intra-node communication. The key distinctions between related works and our work are summarized in Table~\ref{tab:related_works_summary}.

% \begin{table}[h!]
% \centering
% \caption{Comparison of Related Works in Communication Optimization}
% \label{tab:related_works_summary}
% \begin{tabular}{lcccc}
% \toprule
% \textbf{Method} & \textbf{Scope} & \textbf{Lossless} & \textbf{Modifies Compute} & \textbf{Modifies Framework} \\
% \midrule
% Compression & Intra-node & No & No & No \\
% Overlap & Intra-node & Yes & Yes & Yes \\
% Inter-node Multi-path & Inter-node & Yes & No & No \\
% \design{} & Intra-node & Yes & No & No \\
% \bottomrule
% \end{tabular}
% \end{table}
	\section{Design}\label{sec:design}

As shown in Figure~\ref{fig:design}, our system serves as a communication backend for LLM frameworks, playing a role analogous to NCCL. Its core component, the \textit{Communicator} (Section~\ref{subsec:communicator}), serves as the interface between these frameworks and the physical hardware.

The \textit{Communicator} abstracts diverse hardware interconnects—including NVLink, RDMA NICs, and PCIe/C2C—into a unified resource pool. This abstraction allows frameworks to perform concurrent transfers without needing to manage the complexity of each underlying link.

To prevent slower links from bottlenecking high-speed ones, the \textit{Communicator} employs a dynamic load balancing mechanism (Section~\ref{subsec:two_stage_load_balance}). An \textit{Evaluator} component constantly monitors link performance, providing runtime feedback to a \textit{Load Balancer}. The \textit{Load Balancer} then makes fine-grained, dynamic adjustments to the traffic distribution across the links. 

% 机制：使能多链路，然后
% 拓扑感知，多流水线多path，最佳的datapath去串起整条线
\subsection{Communicator}
\label{subsec:communicator}
During the initialization of \design{}, the \textit{Communicator} first initializes NCCL communicators and NVSHMEM contexts. It then defines the topology for intra-node data exchange, adopting a classic yet efficient ring-based model. Specifically, for the PCIe path, the \textit{Communicator} routes any GPU-to-GPU transfer through a designated host memory buffer, which acts as a transit point.

To manage data flow and hide PCIe latency as discussed in Section~\ref{subsec:challenge}, we implement a double-buffered pipeline that decouples data transfer into Producer-Device-to-Host (PD2H) and Host-to-Consumer-Device (H2CD) stages. Dedicating a pinned-memory buffer to each stage allows the PD2H copy of one chunk to overlap with the H2CD copy of another, maximizing PCIe bus utilization. Despite this optimization, we observe that cache misses during memory access lead to inconsistent data transfer rates. To mitigate this, we implement several key optimizations. We bind CPU processes to the physical cores on the NUMA node closest to the GPU to reduce CPU overhead. Furthermore, we allocate the shared pinned-memory buffer in a NUMA-aware manner to fully leverage memory bandwidth and cache performance. Similar NUMA-aware optimizations are applied to the RDMA path.

In ~\design{}, each shared buffer is written by the producer GPU and read by the consumer GPU. We avoid costly memory fences or CPU locks for low-latency synchronization. Instead, we use CUDA's stream-ordered memory operations (\texttt{cuStreamWaitValue32} and \texttt{cuStreamWriteValue32}), which enable GPUs to poll a memory location directly, minimizing CPU involvement and overhead. Binary semaphores are inadequate when a single shared buffer is reused across multiple iterations, because a late write may satisfy a future wait and cause the consumer to read stale data. To ensure correctness, we use a monotonically increasing counter. For an iteration i, the producer waits for \texttt{semEmpty==i}, writes data, and then sets the peer's \texttt{semFull} to \texttt{i+1}. The consumer waits for \texttt{semFull==i+1}, reads the data, and finally sets \texttt{semEmpty} to \texttt{i+1}. This strict ordering prevents stale reads across iterations.

% 策略2：两阶段负载均衡
\subsection{Two-Stage Load Balancing Strategy}
\label{subsec:two_stage_load_balance}
Aggregating heterogeneous communication links is challenging because the overall communication time is dictated by the slowest link. This presents a critical question: how should the communication load be distributed across slower auxiliary paths like PCIe and RDMA? If too much data is offloaded, their higher latency creates a new bottleneck, slowing down the entire operation. Conversely, if too little is offloaded, the performance gain becomes negligible, and the system effectively reverts to an NVLink-only baseline, rendering the multi-path design ineffective.

To address this, we introduce a two-stage load balancing strategy. The approach is to be conservative initially and adaptive at runtime. The first stage establishes a safe and efficient static load distribution through a coarse-grained initial tuning. The second stage implements a fine-grained runtime adjustment mechanism that adapts to dynamic factors like varying message sizes. This approach ensures maximum utilization of all available bandwidth without penalizing the primary NVLink path.

\subsubsection{Stage 1: Initial Coarse-Grained Tuning}

Upon initialization, we perform a brief, one-time profiling phase (approximately 10 s) to find a near-optimal static distribution of communication shares. The goal is a balanced state where all links complete their data transfers in roughly the same amount of time.

The procedure is detailed in Algorithm~\ref{alg:initial_tuning}. The algorithm iteratively measures path completion times to find the slowest and fastest active links. A key principle is our NVLink-centric logic: if NVLink is not the slowest path, we transfer a share of the load from the current slowest path to it. Conversely, if NVLink is the bottleneck, we offload some of its share to the fastest alternative. To ensure stability, the adjustment step size is halved whenever the bottleneck path changes. This acts as a damping factor to prevent oscillation, where the bottleneck might otherwise rapidly shift back and forth between two links. If a path's share is reduced to zero, it is marked as inactive and excluded from subsequent balancing. The process terminates when the timing imbalance falls below a convergence threshold for several iterations, or if NVLink becomes the sole active path, rendering the slower paths ineffective and thus deactivating them.

\subsubsection{Stage 2: Runtime Fine-Grained Adjustment}
While initial tuning provides a robust baseline, the optimal load distribution can vary with data size. A lightweight runtime mechanism handles this dynamism without significant overhead.

\begin{figure}
    \centering
/    \includegraphics[width=1.0\linewidth]{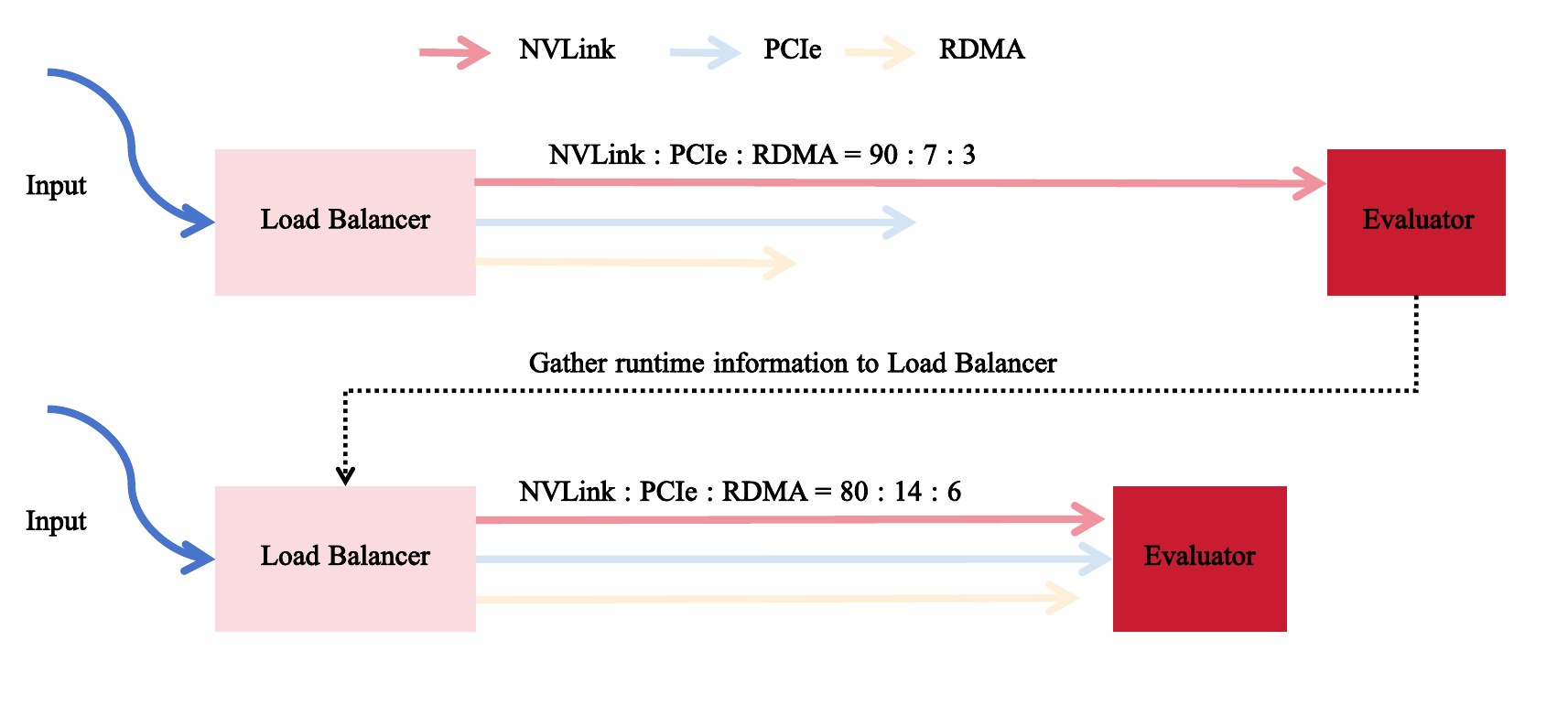}
    \caption{\design{} dynamically adjusts the load based on monitored runtime metrics.}
    \label{fig:load_balance_example}
\end{figure}

This stage employs an \textit{Evaluator} to passively monitor path completion times and a \textit{Load Balancer} to make adjustments. To minimize the overhead from inter-process coordination, the \textit{Load Balancer} is invoked only periodically. \textit{Evaluator} analyzes timings from a recent window (e.g., the last 10 collective calls) of operations to identify persistent trends. If the timing gap between the slowest and fastest paths exceeds a threshold, a small, fixed-size share is transferred from the slowest path to the fastest, prioritizing NVLink. An example is shown in Figure~\ref{fig:load_balance_example}. This gradual approach avoids reacting to transient spikes, ensuring stable convergence with negligible overhead. 

\begin{algorithm}[H]
\caption{Initial Coarse-Grained Load Tuning}
\label{alg:initial_tuning}
\begin{algorithmic}[1]
\State \textbf{Input:} Set of communication paths $C$
\State \textbf{Output:} Converged share distribution $S$
\Statex
\Function{InitialTune}{$C$}
    \State $C_{active} \gets C$
    \State $S \gets \text{InitializeShares}(C_{active})$ \Comment{Heuristic: NVLink gets dominant share}
    \State $step \gets \text{INITIAL\_ADJUSTMENT\_STEP}$
    \State $stability\_count \gets 0$
    \State $prev\_slowest \gets \text{null}$
    \For{$i \gets 1$ to $100$}
        \If{$|C_{active}| = 1$ \textbf{and} $\text{NVLink} \in C_{active}$}
            \State \textbf{break} \Comment{Exit if only NVLink remains}
        \EndIf
        \State $T \gets \text{MeasurePathTimings}(S, C_{active})$
        \State $c_{slow}, c_{fast} \gets \text{FindSlowestFastestPaths}(T, C_{active})$
        \State $imbalance \gets (T[c_{slow}] - T[c_{fast}]) / T[c_{fast}]$
        
        \If{$imbalance < \text{CONVERGENCE\_THRESHOLD}$}
            \State $stability\_count \gets stability\_count + 1$
            \If{$stability\_count \geq \text{STABILITY\_REQUIRED}$}
                \State \textbf{break} \Comment{System is stable}
            \EndIf
        \Else
            \State $stability\_count \gets 0$
            \If{$c_{slow} \neq prev\_slowest$ \textbf{and} $prev\_slowest \neq \text{null}$}
                 \State $step \gets \max(step / 2, 1)$ \Comment{Halve step on bottleneck shift}
            \EndIf
            
            \State $c_{source} \gets c_{slow}$
            \If{$c_{slow} \neq \text{NVLink}$ \textbf{and} $\text{NVLink} \in C_{active}$}
                \State $c_{target} \gets \text{NVLink}$ \Comment{Favor NVLink to maximize its usage}
            \Else
                \State $c_{target} \gets c_{fast}$ \Comment{Offload from bottlenecked NVLink}
            \EndIf
            
            \State $move\_amount \gets \min(step, S[c_{source}])$
            \State $S[c_{source}] \gets S[c_{source}] - move\_amount$
            \State $S[c_{target}] \gets S[c_{target}] + move\_amount$
            
            \If{$S[c_{source}] \leq 0$}
                \State $C_{active} \gets C_{active} \setminus \{c_{source}\}$ \Comment{Deactivate path}
            \EndIf
            \State $prev\_slowest \gets c_{slow}$
        \EndIf
    \EndFor
    \State \textbf{return} $S$
\EndFunction
\end{algorithmic}
\end{algorithm}

\section{Implementation}
The \design{} framework is implemented with approximately 500 lines of Python for orchestration and 3,500 lines of C++/CUDA for the core communication logic. Furthermore, \design{} calls NCCL's APIs for efficient intra-node communication over NVLink and employs NVSHMEM's CPU-initiated APIs to manage communication through the RDMA-capable NICs.

	\section{Evaluation}
\label{sec:evaluation}

\subsection{System and Workload Configuration}
We evaluate \design{} on a server equipped with eight H800 GPUs. The hardware configuration employs a PCIe 5.0 interconnect with an x16 interface, supporting both GPU-to-CPU and inter-GPU communication. This setup provides a theoretical unidirectional I/O bandwidth of 64 GB/s (128 GB/s bidirectional). Each GPU is paired with a dedicated Mellanox ConnectX-6 50 GB/s NIC, connected via a shared PCIe switch. Buffer size affects data transfer speed and efficiency. We empirically select a 4MB buffer for both PCIe and the RDMA paths in our design. The current implementation of \design{} supports and is evaluated on the AllReduce and AllGather collective operations.

\subsection{Baselines and Metrics}
We compare the performance of \design{} with NCCL 2.27.3~\cite{ncclgithub}. NCCL is a state-of-the-art library offering highly optimized implementations of collective communication primitives, primarily designed to maximize throughput. To ensure a fair comparison, we refer to nccl-tests~\cite{nccltestgithub} and report the algorithm bandwidth as our primary performance metric.

\subsection{End-to-End Performance}
\label{subsec:end-to-end_performance}

% Note: Remember to add the booktabs and multirow packages to your main .tex file: 
% \usepackage{booktabs}
% \usepackage{multirow}
\begin{table*}[htbp]
\centering
\footnotesize
\setlength{\tabcolsep}{4.5pt} % Adjusted column spacing
\caption{End-to-end effective algorithm bandwidth (GB/s) and load distribution across various message sizes. The performance of \design{} with both PCIe-only and PCIe+RDMA offloading is compared against the NCCL baseline.}
\label{tab:end_to_end_performance}
\begin{tabular}{lccccccccccc}
\toprule
\multirow{2}{*}{\textbf{Operator}} & \multirow{2}{*}{\textbf{\# GPUs}} & \textbf{Message} & \textbf{NCCL} & \multicolumn{3}{c}{\textbf{\design{} (PCIe-Only)}} & \multicolumn{3}{c}{\textbf{\design{} (PCIe+RDMA)}} \\
\cmidrule(lr){5-7} \cmidrule(lr){8-10}
& & \textbf{Size} & \textbf{Baseline} & \textbf{Total BW} & \textbf{Impr.} & \textbf{PCIe Load} & \textbf{Total BW} & \textbf{Impr.} & \textbf{PCIe + RDMA Load} \\
& & & \textbf{(GB/s)} & \textbf{(GB/s)} & \textbf{(\%)} & \textbf{(\%)} & \textbf{(GB/s)} & \textbf{(\%)} & \textbf{(\%)} \\
\midrule
\multirow{9}{*}{AllReduce} & \multirow{4}{*}{2}
& 32MB & 112 & 131 & 17\% & 14\% & 134 & 20\% & 16 + 4 \\
& & 64MB & 128 & 144 & 13\% & 17\% & 150 & 17\% & 13 + 5 \\
& & 128MB & 132 & 155 & 17\% & 17\% & 165 & 25\% & 11 + 9 \\
& & 256MB & 139 & 167 & 20\% & 18\% & 175 & 26\% & 12 + 9 \\
\cmidrule(lr){2-10}
& \multirow{4}{*}{4}
& 32MB & 87 & 87 & 0\% & 0\% & 89 & 2\% & 2 + 1 \\
& & 64MB & 90 & 97 & 8\% & 8\% & 99 & 10\% & 6 + 2 \\
& & 128MB & 94 & 106 & 13\% & 12\% & 110 & 17\% & 12 + 2 \\
& & 256MB & 98 & 116 & 18\% & 17\% & 118 & 20\% & 13 + 5 \\
\cmidrule(lr){2-10}
& 8
& 256MB & 107 & 108 & 1\% & 1\% & 109 & 2\% & 1 + 1 \\
\midrule
\midrule
\multirow{12}{*}{AllGather} & \multirow{4}{*}{2}
& 32MB & 103 & 122 & 18\% & 15\% & 126 & 22\% & 10 + 8 \\
& & 64MB & 117 & 136 & 16\% & 19\% & 141 & 21\% & 9 + 10 \\
& & 128MB & 129 & 153 & 19\% & 21\% & 153 & 19\% & 12 + 8 \\
& & 256MB & 132 & 163 & 23\% & 21\% & 161 & 22\% & 14 + 5 \\
\cmidrule(lr){2-10}
& \multirow{4}{*}{4}
& 32MB & 43 & 50 & 16\% & 13\% & 52 & 21\% & 10 + 7 \\
& & 64MB & 46 & 56 & 22\% & 18\% & 57 & 24\% & 12 + 8 \\
& & 128MB & 48 & 58 & 21\% & 18\% & 60 & 25\% & 12 + 10 \\
& & 256MB & 49 & 60 & 22\% & 18\% & 62 & 27\% & 12 + 10 \\
\cmidrule(lr){2-10}
& \multirow{4}{*}{8}
& 32MB & 20 & 23 & 15\% & 12\% & 24 & 20\% & 12 + 4 \\
& & 64MB & 21 & 24 & 14\% & 13\% & 26 & 24\% & 12 + 6 \\
& & 128MB & 21 & 25 & 19\% & 14\% & 25 & 19\% & 12 + 7 \\
& & 256MB & 21 & 25 & 19\% & 13\% & 26 & 24\% & 12 + 7 \\
\bottomrule
\multicolumn{11}{l}{\footnotesize{Impr. = Improvement vs. Baseline. The "PCIe + RDMA Load" column shows the respective percentage load on each path.}} \\
\end{tabular}
\end{table*}

As detailed in Table~\ref{tab:end_to_end_performance}, \design{} significantly improves performance by dynamically aggregating bandwidth from NVLink, PCIe, and RDMA-capable NICs.

Our approach effectively offloads traffic to the PCIe and RDMA paths, thereby alleviating congestion on the primary NVLink path. This strategy is particularly effective for AllGather operations, where \design{} consistently improves bandwidth, with gains up to 27\%. The data shows that a significant portion of traffic is diverted; for instance, the PCIe path typically handles 10–14\% of the communication load, while the RDMA path contributes an additional 4–10\%. The consistent load carried by the RDMA path demonstrates a clear performance advantage compared to a PCIe-only offloading strategy, validating the inclusion of the network interconnect in our multi-path design.

For AllReduce operations, \design{} also demonstrates notable gains, especially in 2-GPU and 4-GPU configurations. However, the improvement becomes marginal in the 8-GPU AllReduce scenario. This is attributed to the high latency sensitivity of its underlying ring-based algorithm. A Ring AllReduce requires $2(N-1)$ sequential steps, which is double the $N-1$ steps of AllGather. For an 8-GPU setup ($N=8$), the high latency of the PCIe and RDMA paths is amplified across 14 communication steps, creating a prohibitive bottleneck. This cumulative latency penalty outweighs the benefits of bandwidth offloading. Consequently, our scheduler correctly limits traffic diversion in this specific case to avoid performance degradation, exploiting almost entirely the low-latency NVLink fabric.

\subsection{Overhead Analysis}
\label{subsec:overhead}
Our multi-path approach introduces modest resource overhead. First, managing the PCIe and RDMA data paths incurs some CPU overhead for polling and transfer coordination. This overhead is embedded within the RDMA and DMA control flows, so provided the latency is not excessive, it at worst results in performance comparable to NCCL, rather than a net loss. Second, efficient DMA transfers necessitate the use of pinned host memory buffers; in our configuration, we allocate 4 MB for each path. This pinned buffer requirement consumes a portion of host memory. Finally, the coordination kernels introduce a certain amount of Streaming Multiprocessor (SM) resources, which we plan to analyze and optimize in future iterations. These overheads are generally minimal and are outweighed by the significant bandwidth gains in most scenarios.

% \subsection{Breakdown and Sensitivity Analysis}

	\section{Limitations and Future Work}
Our work has several limitations. First, the current RDMA implementation, which relies on the NVSHMEM CPU API, is suboptimal and requires further optimization. Second, our method introduces certain overhead on the SMs that needs to be minimized. Finally, the effectiveness of \design{} is contingent on the availability of PCIe bandwidth; its performance benefits may be diminished when the PCIe bus is heavily used by other designs~\cite{xu2024piepoolingcpumemory, he2024fastdecodehighthroughputgpuefficientllm}.

For future work, we plan to extend \design{} to support a broader range of communication primitives, such as AllToAll. To further optimize the 8-GPU AllReduce latency, we will explore alternatives like tree-based algorithms and increasing the pipeline depth for the ReduceScatter part to reduce potential bubbles caused by reduce sum computation. We also intend to integrate our solution into major deep learning frameworks, including Megatron-LM~\cite{shoeybi2020megatronlmtrainingmultibillionparameter}, SGLang~\cite{zheng2024sglang}, and vLLM~\cite{kwon2023vllm}, for comprehensive end-to-end evaluation. Furthermore, we aim to extend its applicability to other hardware platforms with constrained interconnect bandwidth, as we anticipate such hardware will continue to play a significant role in the ecosystem.

Looking ahead, emerging interconnect architectures, such as the GB300-generation GPU server, will eliminate bandwidth contention mentioned in observation~\ref{subsec:link_idleness}, further unlocking the potential of multi-link communication strategies.

\section{Conclusion}
We presented \design{}, a library that mitigates communication bottlenecks in large models by dynamically aggregating NVLink, PCIe, and RDMA links into a unified pool. \design{} uses a two-stage adaptive load balancing mechanism to manage traffic, preventing slower links from stalling faster ones. On H800 GPUs, \design{} improves AllReduce and AllGather bandwidth by up to 26\% and 27\% over NCCL respectively. Its full NCCL API compatibility makes \design{} a lossless and easy-to-adopt drop-in solution.

	\bibliographystyle{plainnat}
	\bibliography{btr}
\end{document}